\documentclass[]{elsart}
\usepackage{amsmath, amsfonts, latexsym}
\usepackage{amssymb}
\usepackage{graphicx}
\usepackage{epsfig}

\begin{document}

\begin{frontmatter} 

\title{
Atmospheric effects on extensive air showers observed 
with the Surface Detector of the Pierre Auger Observatory
}


\par\noindent
{\bf The Pierre Auger Collaboration} \\
J.~Abraham$^{8}$, 
P.~Abreu$^{71}$, 
M.~Aglietta$^{54}$, 
C.~Aguirre$^{12}$, 
E.J.~Ahn$^{87}$, 
D.~Allard$^{31}$, 
I.~Allekotte$^{1}$, 
J.~Allen$^{90}$, 
P.~Allison$^{92}$, 
J.~Alvarez-Mu\~{n}iz$^{78}$
% ...

\par\noindent
$^{1}$ Centro At\'{o}mico Bariloche and Instituto Balseiro (CNEA-
UNCuyo-CONICET), San Carlos de Bariloche, Argentina \\
% ...
$^{8}$ Observatorio Meteorologico Parque Gral.\ San Martin (UTN-
FRM/CONICET/CNEA), Mendoza, Argentina \\
% ...
$^{12}$ Universidad Catolica de Bolivia, La Paz, Bolivia \\
% ...
$^{31}$ Laboratoire AstroParticule et Cosmologie (APC), 
Universit\'{e} Paris 7, CNRS-IN2P3, Paris, France \\
% ...
$^{54}$ Istituto di Fisica dello Spazio Interplanetario (INAF),
 Universit\`{a} di Torino and Sezione INFN, Torino, Italy \\
% ...
$^{71}$ LIP and Instituto Superior T\'{e}cnico, Lisboa, Portugal \\
% ...
$^{78}$ Universidad de Santiago de Compostela, Spain \\
% ...
$^{87}$ Fermilab, Batavia, IL, USA \\
% ...
$^{90}$ New York University, New York, NY, USA \\
% ...
$^{92}$ Ohio State University, Columbus, OH, USA \\
% ...
\par\noindent
(\ddag) Deceased \\
(a) at Konan University, Kobe, Japan \\
(b) On leave of absence at the Instituto Nacional de Astrofisica, Optica y Electronica \\
(c) at Caltech, Pasadena, USA \\

\begin{abstract}
Atmospheric parameters, such as pressure ($P$), temperature ($T$) and
density ($\rho \propto P/T$), affect the development of extensive air
showers initiated by energetic cosmic rays. We have studied the
impact of atmospheric variations on extensive air showers
by means of the surface detector of the Pierre Auger Observatory.
The rate of events shows a
$\sim 10\%$ seasonal modulation and $\sim 2\%$ diurnal one. We find
that the observed behaviour is explained by a model including the effects
associated with the variations of $P$ and
$\rho$. The former affects the longitudinal development of air showers
while the latter influences the Moli\`ere radius and hence the lateral
distribution of the shower particles. The model is validated with full
simulations of extensive air showers
using atmospheric profiles measured at the site of the Pierre Auger Observatory.

\end{abstract}

\begin{keyword}
extensive air showers \sep UHECR \sep atmosphere \sep weather
\PACS 96.50.sd \sep 96.50.sb \sep 96.50.sf

\end{keyword}

\end{frontmatter}  


%
\section{Introduction}
\label{sec:introduction}

High-energy cosmic rays (CRs) are measured by recording the extensive air
showers (EAS) of secondary particles they produce in the atmosphere. As the
atmosphere is the medium in which the shower evolves, its state affects the
lateral and longitudinal development of the shower. 
Pressure ($P$) and air density ($\rho$) are the properties of the atmosphere
that mostly affect the EAS. An increase (or decrease) of the
ground $P$ corresponds to an increased (or decreased) amount of matter 
traversed by the shower particles; 
this affects the stage of the longitudinal development of the shower when 
it reaches the ground.
A decrease (or increase) of $\rho$ increases (or decreases) the Moli\`ere radius and
thus broadens (or narrows) the lateral extent of the EAS. 

The properties of the
primary CR, e.g.,  energy, mass and arrival direction, have to be inferred from
EAS, which can be sampled by an array of detectors at ground level. 
Therefore the study and understanding of the effects of atmospheric variations
on EAS in general, and on a specific detector in particular, is very important for
 the comprehension of the
detector performances and for the correct interpretation of EAS measurements. 

We have studied the atmospheric effects on EAS by means of the surface detector
(SD) of the Pierre Auger Observatory, located in Malarg\"{u}e, Argentina
(35.2$^\circ$ S, 69.5$^\circ$W) at 1400 m a.s.l. \cite{auger}.
The Pierre Auger Observatory is designed to study CRs from $\sim 10^{18}$~eV
up to the highest energies. The SD consists of 1600 water-Cherenkov detectors to
detect the photons and the charged particles of the showers. It is laid out over
3000~km$^{2}$ on a triangular grid of 1.5~km spacing \cite{augerSD} and is
overlooked by four fluorescence detectors (FD) \cite{augerFD}. The SD trigger
condition, based on a 3-station coincidence \cite{augerTrigger}, makes the
array fully efficient above about $3 \times 10^{18}$~eV. For each event, the
signals in the stations  are fitted to find the signal at 1000~m from the shower
core, $S(1000)$, which is used to estimate the primary energy
\cite{newton}. The atmosphere is continuously monitored by different
meteorological stations located at the central part of the array and at each FD
site. In addition, balloon-borne sensors are launched at regular intervals to
measure the atmospheric temperature $T(h)$, pressure $P(h)$ and
humidity $u(h)$ as a function of the altitude $h$ above the detector
\cite{keilhauer}.

In section~\ref{sec:model}, we develop a model of the expected atmospheric
effects on $S(1000)$. The modulation is described by means of three coefficients
that depend on the EAS zenith angle ($\theta$). They are related to variations
of $P$ and $\rho$, measured at ground level, 
on \emph{slower} (daily-averaged) and
\emph{faster} (within a day) time scales. 
The dependence of $S(1000)$ on $P$ and
$\rho$ implies a modulation of the counting rate of events.
In section~\ref{sec:rate}, we study the behaviour of the recorded rate of events
as a function of $P$ and $\rho$. On the base of the model defined previously, we
derive the $P$ and $\rho$ coefficients.  In section~\ref{sec:simulations}, we
perform full simulations of EAS developing in various realistic atmospheres
(based on measurements from balloon soundings above the site of the Pierre Auger
Observatory)  in order to compare, in section~\ref{sec:comparison}, the results
from data and simulations with the predictions of the model. We conclude in
section~\ref{sec:conclusion}.

%
\section{Model of atmospheric effects for the surface detector of the Auger Observatory}
\label{sec:model}
\subsection{Atmospheric effects on the measured signal}
The water-Cherenkov detectors are sensitive to both the electromagnetic 
component and the muonic component of the EAS,  which are influenced to a
different extent by atmospheric effects, namely by variations of $P$ and $\rho$.
These in turn influence the signal measured in the detectors: for the Auger
Observatory,  we are in particular interested in the effects on the signal at
1000~m from the core, $S(1000)$.

The continuous measurement of atmospheric $P$ and $\rho$ is available only
at ground level. We will show that the variation of $S(1000)$ can be fully
described in terms of variation of air pressure and air density measured at
the altitude of the Observatory site.
If not otherwise stated, $P$ and $\rho$ refer to the values at ground level.

In the following, we first describe separately the effects on $S(1000)$ due to
$P$, section~\ref{sec:pressure},  and $\rho$, section~ \ref{sec:density}, 
and then in section \ref{sec:total} we provide the full 
parameterisation of its variations
as a function of changes in $P$ and $\rho$.

\subsubsection{Effect of air pressure variations on the SD signal} 
\label{sec:pressure}
From the point of view of $P$ (which measures the vertical air column
density above ground), an increase (decrease) corresponds to an increased
(decreased) matter overburden. This implies that the shower is older (younger),
i.e. in a more (less) advanced stage when it reaches the ground level.

\begin{figure}
\begin{center}
\centerline{
\includegraphics[draft = false,scale = 0.45]{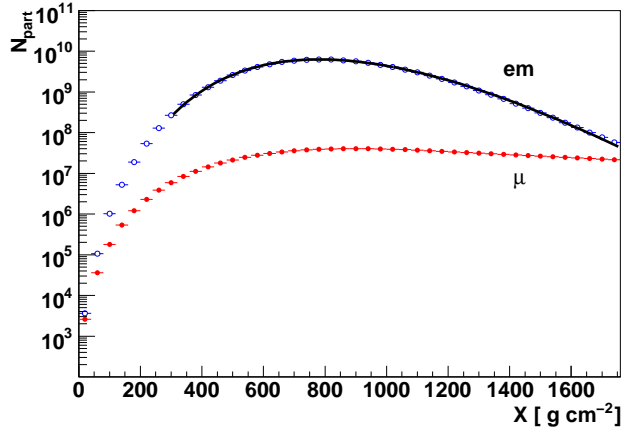}}
\caption{Average longitudinal profile of three hundred  proton-initiated showers
with E~=~$10^{19}$~eV, and zenith angle $\theta = 60^{\circ}$, 
simulated with CORSIKA-QGSJETII
(open blue circles represent the electromagnetic component, red bullets the muonic one).
The black continuous line is a fit of the electromagnetic profile with a Gaisser-Hillas
function.}
\label{fig:longProf}
\end{center}
\end{figure}

The longitudinal profile of the electromagnetic component of the EAS is exponentially
attenuated beyond the shower maximum and can be described by a Gaisser-Hillas
profile \cite{GH} (see Fig.~\ref{fig:longProf}).
We are interested in the value of the electromagnetic signal measured at 1000~m from the
core, referred hereafter as $S_{em}$. The longitudinal development of the shower
far from the core is delayed with respect to the one at the core, and can be
parameterised as
\begin{equation}
\nonumber
S_{em}(E,X)\propto X^{\hat X_{max}/\Lambda}\exp[(\hat X_{max}-X)/\Lambda],
\end{equation}
where $E$ is the primary energy, $X$ the slant depth, $\hat X_{max}\equiv
X_{max}+\Delta$ the average maximum of the shower at 1000~m from the core with 
$X_{max}$
being the shower maximum\footnote{$X_{max}\simeq 750$~g~cm$^{-2}$ for $10^{19}$~eV
showers according to the elongation rate measurement with the FD at the Pierre
Auger Observatory \cite{augerER}},  $\Delta \simeq$~150~g cm$^{-2}$ is the
typical increase of the shower maximum at 1000~m from the core \cite{billoir}
and $\Lambda\simeq 100$~g cm$^{-2}$ is the effective attenuation length after
the maximum \cite{fabian}.  
Therefore, a change in $P$ affects $S_{em}$:
\begin{equation}
\label{eq:dsdp}
\frac{1}{S_{em}}\frac{{\rm d}S_{em}}{{\rm d}P} 
\simeq 
-\frac{1}{g}\left[1-\frac{\hat X_{max}}{X}\right]\frac{{\rm sec}\,\theta}{\Lambda}
\end{equation}
where $g$~d$X={\rm d}P\,{\rm sec}\,\theta$ is used, with $g$ the acceleration of
gravity, and $\theta$ the shower zenith angle. 
Due to the flat longitudinal development of the muons (see
Fig.~\ref{fig:longProf}), no significant pressure dependence is expected for the
muonic component.

\subsubsection{Effect of air density variations on the SD signal} 
\label{sec:density}
Regarding $\rho$, this affects the Moli\`ere radius $r_M$
\begin{equation}
\nonumber
r_M\equiv \frac{E_s}{E_c}\frac{X_0}{\rho}\simeq 
\frac{91\ {\rm m}}{ \rho / ({\rm kg~m}^{-3})}
\end{equation} 
where $E_s\equiv m_ec^2\sqrt{4\pi/\alpha} \simeq 21$~MeV is the energy constant
characterising the energy loss due to multiple Coulomb scattering,  $E_c\simeq
86$~MeV is the critical energy in air and $X_0\simeq 37.1$~g~cm$^{-2}$ is the
radiation length in air. 
A variation in $r_M$ affects the lateral distribution of the electromagnetic component of
the EAS, which can be approximately described with a Nishimura-Kamata-Greisen
(NKG) profile \cite{NKG, greisen}. 
At a large distance $r$ from the core, it behaves as $S_{em}(r)\propto
N_{em}(r)\propto r_M^{-2}(r/r_M)^{-\eta}$, where $\eta\simeq 6.5-2s$ and $s =
3X/(X+2X_{max})$ is the age of the shower. Hence, a change in $\rho$ affects
$S_{em}$:
\begin{equation}
\frac{1}{S_{em}}\frac{{\rm d}S_{em}}{{\rm d}\rho} 
\simeq \frac{(2-\eta)}{\rho} .
\label{eq:dsdr}
\end{equation}
In fact, the relevant value of $r_M$ is the one corresponding to the air density
$\rho^{*}$ two radiation lengths above ground \cite{greisen} in the direction of
the incoming shower. 
This corresponds to $\simeq 700\ {\rm m}\cos\theta$ above the site of the Pierre
Auger Observatory.  On time scales of one day or more, the temperature gradient
(d$T$/d$h$) in the lowest layers of the atmosphere (the planetary  boundary
layer, which extends up to about 1~km above ground level) can be described by an
average value of $\simeq -5.5\,^\circ$C~km$^{-1}$  at the site of the Auger
Observatory. 
Therefore the variation of $\rho^*$ on temporal scales of one day essentially
follows that of $\rho$. An additional effect is related to the diurnal
variations of d$T$/d$h$, because during the day the surface of the Earth is
heated by solar radiation, producing a steeper d$T$/d$h$ in the boundary layer.
On the other hand, during the night the surface is cooled by the emission of
long wavelength radiation: d$T$/d$h$ becomes smaller and even $T$ inversions can
be observed before sunrise. 
As a result, the amplitude of the diurnal variation in $T$ (and $\rho$) is
smaller at two radiation lengths above ground than at ground level. It is then
useful to separate the daily modulation from the longer term one introducing the
average daily density $\rho_d$ and the instantaneous departure from it,
$\rho-\rho_d$. 
Therefore, the dependence of $S_{em}$ on $\rho$ can be modeled by
\begin{equation}
\nonumber
S_{em} = S_{em}^0\left[1+\alpha_\rho^{em}(\rho_d-\rho_0)
+\beta_\rho^{em}(\rho-\rho_d)\right]
\end{equation}
where $\rho_0$~=~1.06~kg~m$^{-3}$ is chosen as the reference value of $\rho$
and is the average value measured at the site of the Pierre Auger Observatory
over more than three years (1~Jan~2005 - 31~Aug~2008).

Concerning the muonic component of the signal at 1000~m from the
core, $S_\mu$, its dependence on $\rho$ can be parameterised as
\begin{equation}
\nonumber
S_{\mu} = S_{\mu}^0\left[1+\alpha_\rho^{\mu}(\rho_d-\rho_0)\right].
\end{equation}
The $\rho$ dependence is written in terms of $\rho_d-\rho_0$ only because, as
the muons are produced high in the atmosphere, their  contribution to signal is
not expected to depend on the  daily modulations taking place in the boundary
layer.

\subsubsection{Model of atmospheric effects on S(1000)}
\label{sec:total} 
\begin{figure}
\begin{center}
\centerline{
\includegraphics[draft = false,scale = 0.5]{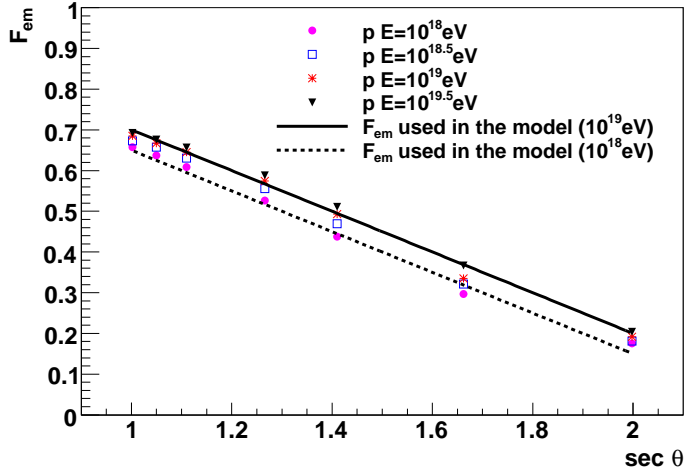}}
\caption{Fraction of the
total signal induced by the electromagnetic component of the shower 
at ground level at a distance of 1000~m from the shower axis ($F_{em}$) 
as a function of ${\rm sec}\,\theta$. 
A linear dependence of $F_{em}$ on ${\rm sec}\,\theta$ (solid and dashed lines)
is assumed in this work.}
\label{fig:Fem}
\end{center}
\end{figure}

The dependence of the total signal at 1000~m from the core, $S(1000)\equiv S
= S_{em}+S_\mu$, upon $P$ and $\rho$ can hence be written as
\begin{equation}
\label{eq:Sgeneral}
S = S_0\left[1+\alpha_P(P-P_0)+\alpha_\rho(\rho_d-\rho_0)
+\beta_\rho(\rho-\rho_d)\right]
\end{equation}
where $P_0 = 862$~hPa is the reference $P$ at  the site of the Pierre Auger
Observatory, $S_0$ is the value of the total signal at reference pressure and
density ($P=P_0$ and $\rho=\rho_d=\rho_0$), and
\begin{equation}
\label{eq:CoeffGeneral}
\alpha_P = F_{em}\alpha_P^{em}
\hspace{1cm}
\alpha_\rho = F_{em}\alpha_\rho^{em}+(1-F_{em})\alpha_\rho^\mu
\hspace{1cm}
\beta_\rho = F_{em}\beta_\rho^{em}
\end{equation}
where $F_{em}\equiv S_{em}/S$ is the electromagnetic fraction of the signal at
1000~m from the core. The values of $F_{em}$ are obtained by means of
proton-initiated showers simulated with CORSIKA-QGSJETII (see
section~\ref{sec:simulations}):  they decrease approximately linearly with
sec~$\theta$  for all the simulated primary energies (see Fig.~\ref{fig:Fem}). 

We will adopt hereafter
\begin{equation}
\label{eq:fem}
F_{em} = F_{em}^v -0.5(\mathrm{sec} \, \theta-1)
\end{equation}
where $F_{em}^v \equiv F_{em}(\theta=0)$ varies between $\approx 0.65$ at
$10^{18}$~eV and $\approx 0.7$ at $10^{19}$~eV. We note that since the inferred
electromagnetic fraction depends on the hadronic model adopted and on the CR
composition assumed, the actual value of $F_{em}$ may be different. 
As shown in \cite{fabian}, for iron-induced showers  the simulated $S_{\mu}$ is
40\% higher than in the case of protons,  while the SIBYLL model \cite{sibyll}
predicts a muonic signal 13\% lower than  QGSJETII for both proton and iron
primaries. 
The corresponding variation $F_{em}^v$ at a primary energy of $10^{19}~eV$ would
be $\simeq -11\%$ for iron with respect to proton, and  $\simeq +4\%$ for SIBYLL
simulations with respect to QGSJETII. 

Finally, with respect to the coefficients in eq.~\ref{eq:CoeffGeneral}:

(i) for the pressure coefficient, we have from eq.~\ref{eq:dsdp}
\begin{equation}
\nonumber
\alpha^{em}_P\simeq 
-\frac{1}{g}\left[1-\frac{\hat X_{max}}{X}\right]\frac{{\rm sec}\,\theta}{\Lambda}
\end{equation}
where $X = X_v \sec\theta$ and $X_v\simeq 880$~g~cm$^{-2}$ is the
atmospheric depth at the site of the Pierre Auger Observatory.

(ii) From eq.~\ref{eq:dsdr}
\begin{equation}
\nonumber
\alpha^{em}_\rho\simeq -\frac{4.5-2s}{\rho}
\end{equation}

where $s = 3/(1+2\cos\theta\ X_{max}/X_v)$, with $X_{max}/X_v\simeq 0.85$ for
$10^{19}$~eV primaries.  
Pressure effects associated to the change in the slope of the lateral
distribution function due to the $X$ dependence of $s$ are negligible.

(iii) The coefficient $\beta_\rho^{em}$ should be smaller than
$\alpha^{em}_\rho$ (in absolute value) reflecting the reduction in the amplitude
of the $\rho-\rho_d$ variations two radiation lengths above ground level. The
difference should also depend on  $\theta$. For instance, assuming an
exponential decrease of the density amplitude with the height $h$
\begin{equation}
\nonumber
\rho(h)-\rho_d(h)=\exp\left(-a\frac{h}{700\ {\rm m}}\right)
[\rho(0)-\rho_d(0)]
\end{equation}
would lead to
\begin{equation}
\label{eq:betvsal}
\beta^{em}_\rho \simeq \exp(-a\cos\theta) \, \alpha^{em}_\rho
\end{equation}

where $a$ parameterises the amplitude of the daily density variation in the
lower atmosphere and is completely independent of the shower development. 
It characterises the scale height for the decrease of the daily thermal 
amplitude, which becomes $1/e$ of its ground value at a height  $(700$~m$)/a$.
The value of $a$ is expected to be of order unity.

(iv) The coefficient $\alpha_\rho^{\mu}$ is expected to be small, and will be
assumed to be independent of $\theta$, because of the relatively flat
longitudinal development of the muons as shown in Fig.~\ref{fig:longProf}. Its
value will be taken to be zero since the air shower simulations are consistent
with a vanishing $\alpha_\rho^{\mu}$ coefficient (see
section~\ref{sec:simulations}).
\subsection{Atmospheric effects on the event rate} 
\label{sec:modelrate}
The dependence of the measured signal on variations of $P$ and $\rho$ produces
also a modulation of the rate of recorded events. The trigger
probability, $P_{tr}$, is a well defined function of the signal
\cite{augerTrigger}. As atmospheric variations correspond to signal variations,
this implies that the same primary particle (in particular,  with the same
primary energy) will induce different signals depending on $P$ and $\rho$. This
in turn affects the probability for the shower to trigger the SD array.

The effect can be quantified starting from the relation between $S(1000)$ and
the energy of the primary cosmic ray. In the case of the Pierre Auger
Observatory, the primary energy is reconstructed as 
\begin{equation}
\nonumber
E_{r} \propto \left[S(1000)\right]^B, 
\end{equation}
where $B = 1.08\pm0.01(stat)\pm0.04(sys)$  is derived
from the calibration of the SD energy using the FD energy measurement \cite{augerSpectrum}. 
Following eq.~\ref{eq:Sgeneral}, the primary energy $E_{0}(\theta,P,\rho)$ that
would have been obtained for the same shower at the reference pressure $P_0$ and
density $\rho_0$, is related to $E_r$ as follows 
\begin{equation}
\label{eq:energy}
E_{0} = E_{r} \left[1-\alpha_P(P-P_0)-\alpha_\rho(\rho_{d}-\rho_0)
-\beta_\rho(\rho-\rho_{d})\right]^B .
\end{equation}
In a zenith angle bin d$\theta$, the rate $R$ of events per unit time and unit
solid angle above a given signal $S_{min}$ can be written as
\begin{equation}
\nonumber
\frac{{\rm d}{R}}{{\rm d}\theta}(\theta,S_{min}) = 
\frac{{\rm d}{A}}{{\rm d}\theta}(\theta)
\int_{S_{min}}{\rm d}S \,P_{tr}(S)\frac{{\rm d}J} {{\rm d}S}
\end{equation}
where $A$ is the geometrical aperture and $J$ is the flux of cosmic rays.

Assuming that the cosmic ray spectrum is a pure power law, i.e. d$J/{\rm
d}E_0\propto E_0^{-\gamma}$, using eq.~\ref{eq:energy}, and neglecting the small
energy dependence of the weather coefficients, we find that
\begin{eqnarray*}
\nonumber
\frac{{\rm d}J}{{\rm d}S} & \propto & E_0^{-\gamma}\frac{{\rm d}E_0}{{\rm d}S}\\
\nonumber
& \propto &  S^{-B\gamma+B-1}\left[1+B(\gamma-1)\left(\alpha_P(P-P_0) + 
\alpha_\rho(\rho_{d}-\rho_0)+\beta_\rho(\rho-\rho_{d})\right)\right] .
\end{eqnarray*}
From the dependence on the atmosphere of the measured CR flux above a given
signal, we derive the corresponding dependence of the rate of events. If $S_{min}$ is
the minimum required signal at 1000~m from the core to trigger the array
\begin{equation}
\label{eq:rate}
\frac{{\rm d}{R}}{{\rm d}\theta}\propto \left[1+a_P(P-P_0) +
a_\rho(\rho_{d}-\rho_0)+b_\rho(\rho-\rho_{d})\right]
\int_{S_{min}}{\rm d}S\
P_{tr}(S) S^{-B\gamma+B-1}
\end{equation}

with the integral on the right hand side being independent of the weather
variations. The coefficients $a_P$, $a_{\rho}$ and $b_{\rho}$ are then related to
the coefficients describing the modulation of the signal by $a_{\rho,P} =
B(\gamma-1)\alpha_{\rho,P}$ and $b_\rho = B(\gamma-1)\beta_{\rho}$.

%
\section{Atmospheric effects on the experimental rate of events}
\label{sec:rate}
To study the modulation of the rate of events, we use data taken by the
SD from 1 January 2005 to 31 August 2008. All events with $\theta < 60^\circ$ are
used, for a total of about 960$\,$000 showers with a median energy
$6~\times~10^{17}$~eV. These are selected on the basis of the topology and time
compatibility of the triggered detectors \cite{augerTrigger}. The station with the
highest signal must be enclosed within an \emph{active hexagon}, in which all six
surrounding detectors were operational at the time of the event. 

\begin{figure}
\begin{center}
\centerline{
\includegraphics[draft = false,scale = 0.7]{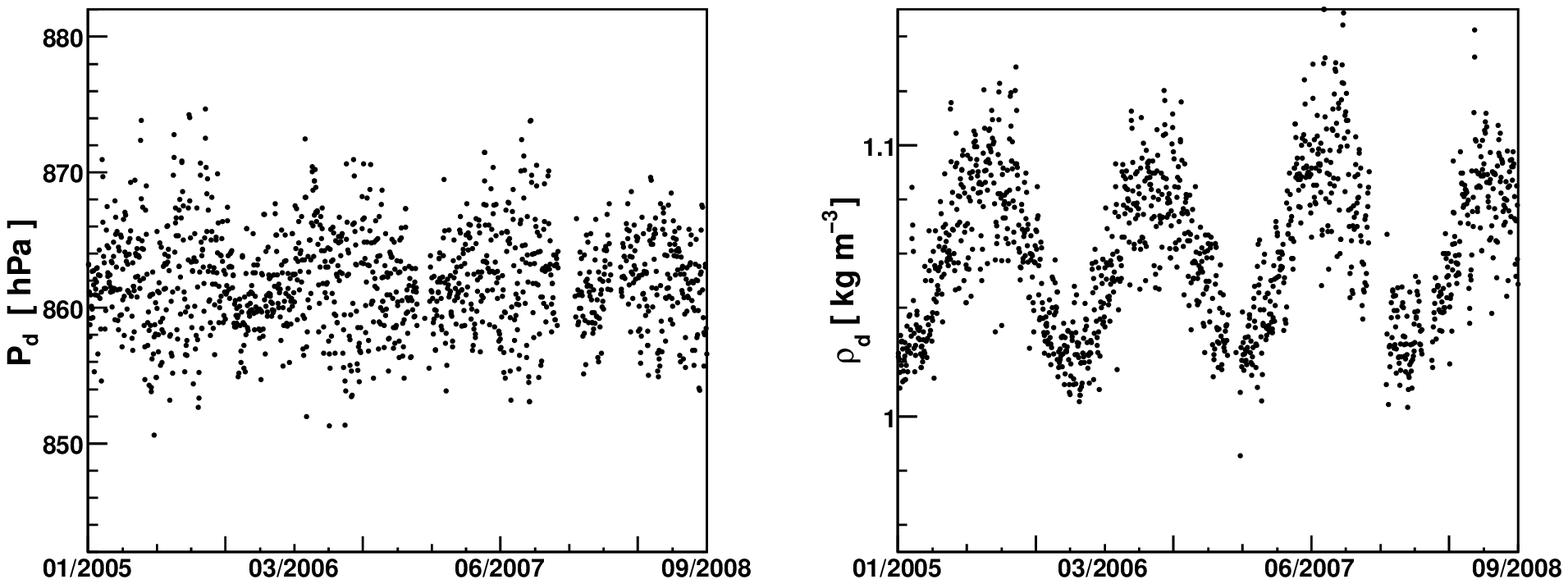}}
\centerline{
\includegraphics[draft = false,scale = 0.7]{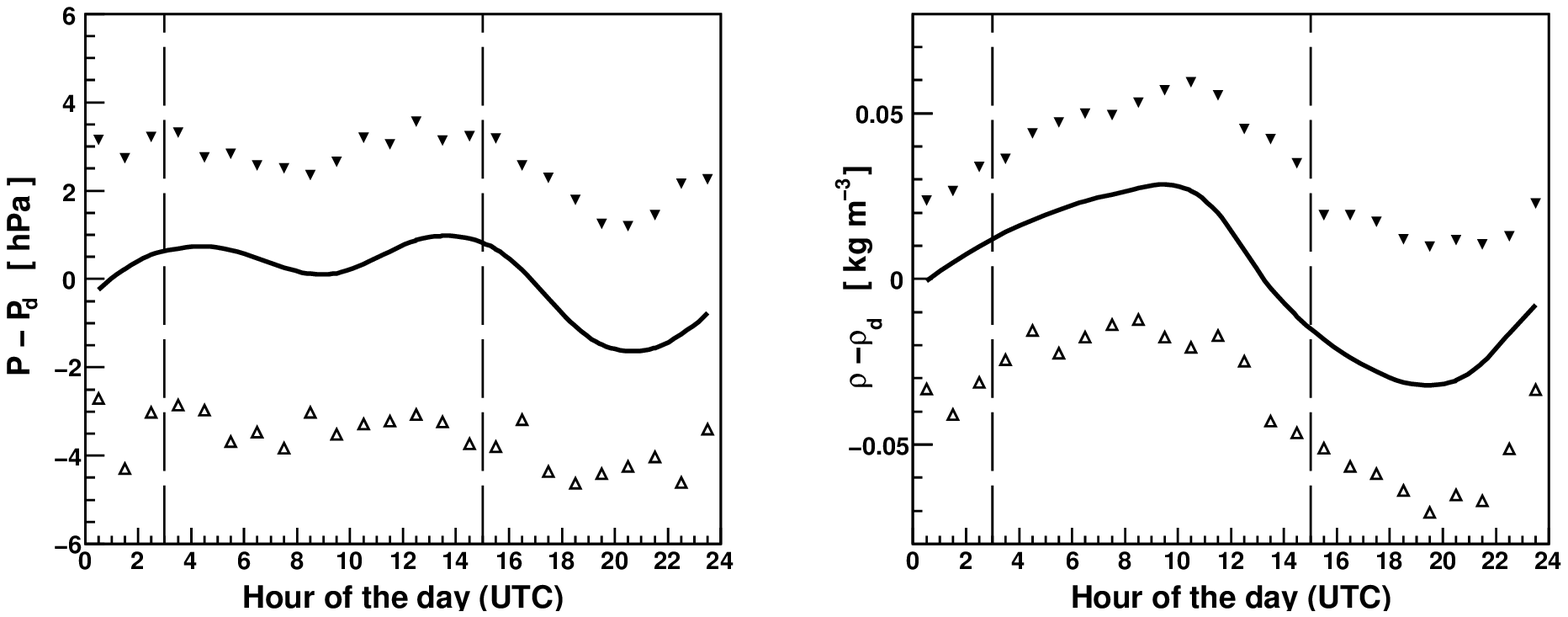}}
\caption{Top: daily averages of $P$ (left) and $\rho$ (right). Bottom: diurnal
variation of $P$ (left) and $\rho$ (right). The values are averaged over the three
years considered (line), with the maximum and minimum variations
marked by black and white triangles. The local time is UTC-3~h (vertical lines mark local midnight
and noon).}
\label{fig:weather}
\end{center}
\end{figure}

At the site of the Pierre Auger Observatory, the ground temperature and pressure
are measured every five minutes.  The air
density is given by: $\rho = ({M_m}/{R})~({P}/{T})$
where $M_m$ is the molecular mass of air, $R$ the gas constant.  
The daily average density $\rho_d$ is
obtained with a smoothing procedure consisting in taking, for each time, the
average value of $\rho$  over a 24~h interval centered at the time of interest.
The daily and diurnal variations of the ground $P$ and $\rho$ are shown in
Fig.~\ref{fig:weather} (upper and lower panels respectively). 
The pressure exhibits less than
$\pm2\%$ variation during the period considered, while $\rho_d$ changes up to a
maximum of $\pm8\%$ with an additional diurnal variation of density which is of
$\pm3\%$ on average with maximum values of $^{+6}_{-8}\%$.

\begin{figure}
\begin{center}
\centerline{
\includegraphics[draft = false,scale = 0.4]{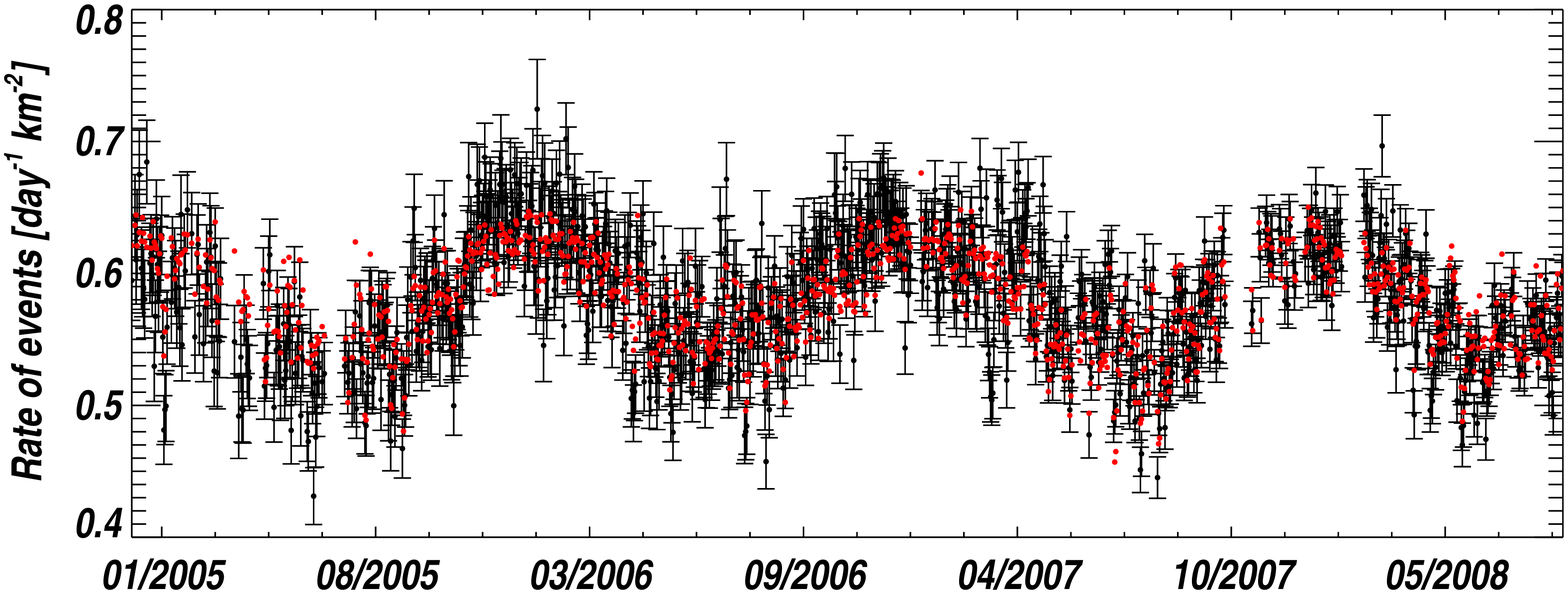}}
\vspace{0.5cm}
\centerline{
\includegraphics[draft = false,scale = 0.4]{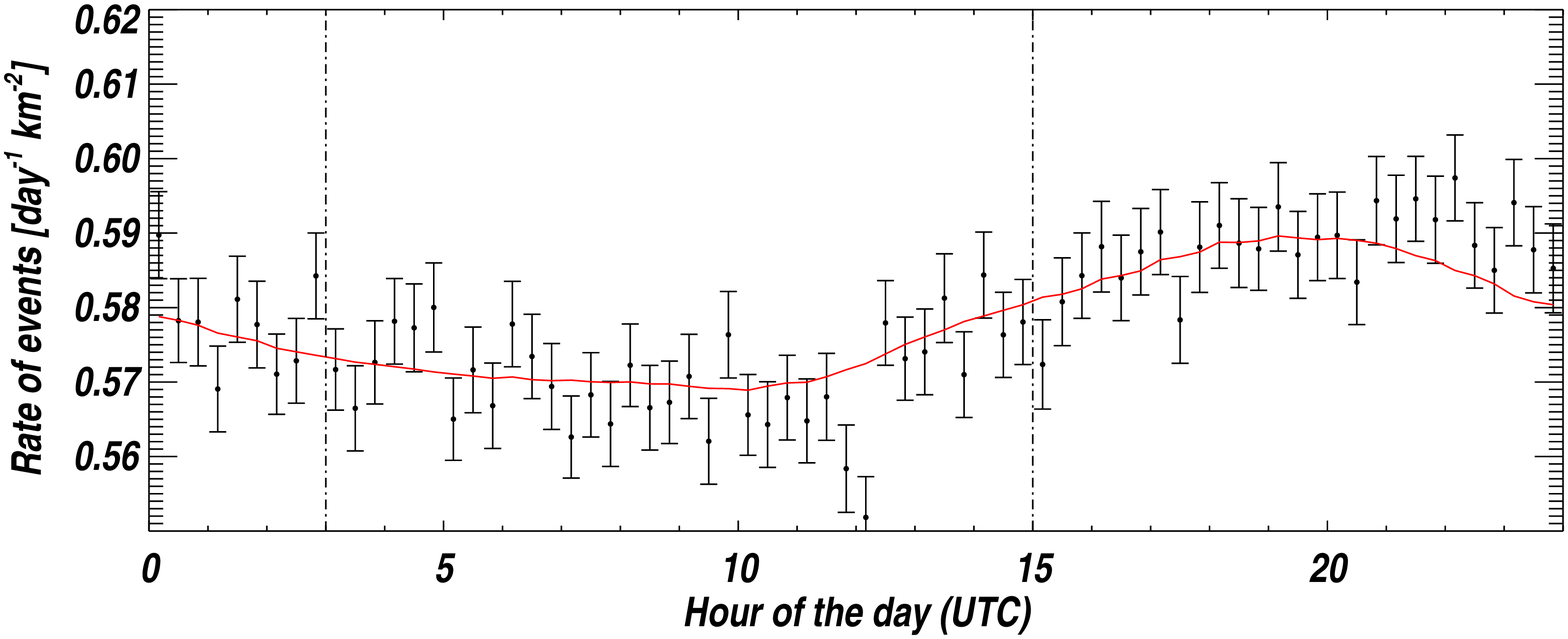}}
\caption{Top: seasonal modulation of the measured
(grey) and fitted (black points) rate of events. Bottom: diurnal
modulation of the measured (grey) and fitted (black line) event rate.}
\label{fig:rate}
\end{center}
\end{figure}

In the period under study, the number of surface detectors steadily increased from
about 700 to about 1590. To take this into account, rather than using the raw
number of triggering events, we compute the rate every hour normalized to the
sensitive area, which is calculated every second from the total area of the active
hexagons.  The daily and the diurnal rate of events are presented in
Fig.~\ref{fig:rate} (black points), where it is evident that they both follow
qualitatively the corresponding modulations of pressure and density from
Fig.~\ref{fig:weather}.

We use the expression given by eq.~\ref{eq:rate} to fit the measured rate of
events. Assuming that the number of events $n_i$ observed in each hour bin $i$
follows a Poisson distribution of average $\mu_i$, a maximum likelihood fit is
performed to estimate the coefficients $a_P$, $a_\rho$ and $b_\rho$.\\
 The
likelihood function is $L = \prod \frac{\mu_i^{n_i}}{n_i!} e^{-\mu_i}$. The
expected number of events in bin $i$ is given by
\begin{equation}
\nonumber
\mu_i = R_0 \times A_i \times C_i
\end{equation}
where $R_0$ is the average rate we would have observed if the atmospheric 
parameters were always the reference ones, i.e. $R_0 = \frac{\sum n_i}{\sum A_i
C_i}$, with $A_i$ the sensitive area in the $i^{th}$ bin and, according to
eq.~\ref{eq:rate}, $C_i$ is 
\begin{equation}
\nonumber
C_i = [1+a_P(P_i-P_0)+a_\rho(\rho_{d_i}-\rho_0)+b_\rho(\rho_i-\rho_{d_i})].
\end{equation}
The fitted parameters are:
\begin{eqnarray}
\label{eq:RateCoefficients}
a_P      & = & (-0.0027 \pm 0.0003)~\mathrm{hPa}^{-1} \nonumber \\
a_{\rho} & = & (-1.99 \pm 0.04)~\mathrm{kg}^{-1}~\mathrm{m}^3 \\
b_{\rho} & = & (-0.53 \pm 0.05)~\mathrm{kg}^{-1}~\mathrm{m}^3 \nonumber
\end{eqnarray}
corresponding to a reduced $\chi^2$ of $1.06$, where $\chi^2 = \sum_i
(n_i-\mu_i)^2/\mu_i$. The result of the fit is shown in
Fig.~\ref{fig:rate}, compared to the daily averaged and the
shorter term modulations of the measured event rate. 

To check the stability of the coefficients with respect to the energy, the same
study has been done for the subset of events with a reconstructed energy above
$10^{18}$~eV, corresponding to $\simeq 20\%$ of the total statistics. The fitted
coefficients are consistent within the fit uncertainties. A more detailed study of the
energy dependence of the coefficients will become feasible in future with increased
statistics.

%
\section{Atmospheric effects on simulated air showers}
\label{sec:simulations}
To complete the study of atmospheric effects, we performed full EAS simulations in
different atmospheric conditions. We simulated proton-initiated  showers using the
CORSIKA code \cite{corsika} with hadronic interaction models QGSJETII
\cite{qgsjet} and Fluka \cite{fluka}. 

\begin{figure}
\begin{center}
\centerline{
\includegraphics[draft = false,scale = 0.65]{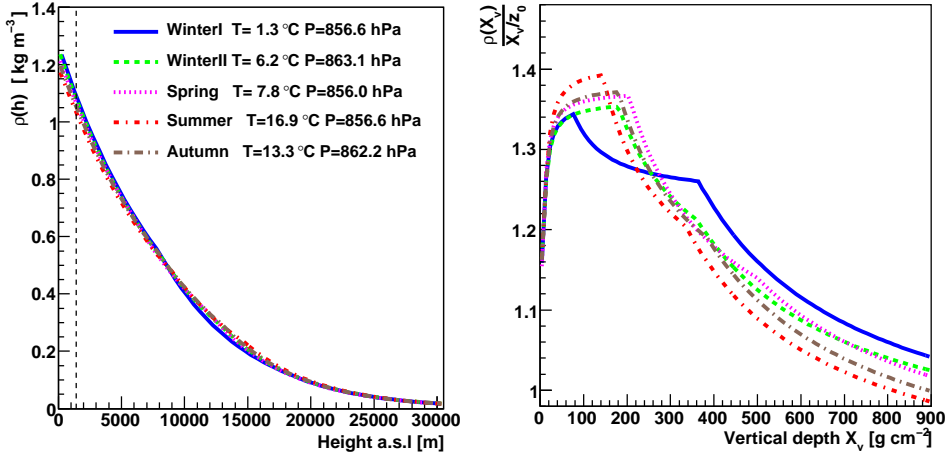}}
\caption{
Left: density profiles used in the simulations. The dashed vertical line
corresponds to the altitude of the Pierre Auger Observatory (1400~m). 
The corresponding values
of ground $P$ and $T$ are given in the legend. Right: same density
profiles normalized to an isothermal one ($\rho(X_v)=X_v/z_{0}$ with
$z_{0}$~=~8.4~km).
}
\label{fig:seasonProf}
\end{center}
\end{figure}

We considered four fixed energies of the primary particle
($E$~=~$10^{18}$~eV,~$10^{18.5}$~eV, $10^{19}$~eV and $10^{19.5}$~eV)  and seven
fixed zenith angles between $\theta = 0^{\circ}$ and $\theta = 60^{\circ}$.  For
the air density profiles, we used five parameterisations (shown in
Fig.~\ref{fig:seasonProf}) of the seasonal average of radio sounding  campaigns
carried out at the site of the Pierre Auger Observatory \cite{keilhauer}  over a
wide range of variation in temperature\footnote{The atmospheric profiles are
implemented in the CORSIKA code through the dependence of $X$ on $h$. $P$, $\rho$
and $T$ profiles can be derived from: $\rho(h) = -{\rm d}X/{\rm d}h$ and $P(h) = g
X(h)$. The ground values in Fig.~\ref{fig:seasonProf} are computed at an
observation level $h = 1400$~m ($\simeq$~880~g cm$^{-2}$), corresponding to the
altitude of the Pierre Auger Observatory.}. 
The set of simulations consists of 60 showers for each combination of atmospheric
profile, energy and angle with an optimal statistical thinning level of $10^{-6}$
\cite{thinning, optimalThinning}.

To compare with model predictions and data, we need to determine for each
combination ($E$,~$\theta$) the dependence of $S(1000)$ on the variations of $P$
and $\rho$. The signal can be estimated through simplified assumptions about the
energy deposited by particles on the basis of their kinetic energy $E_k$:\\
(i) e$^-$e$^+$ deposit $E_k-\epsilon_{th}$, where $\epsilon_{th} =
260$~keV is the energy threshold for Cherenkov emission in water.\\
(ii) photons deposit $E_k-2m_e-2\epsilon_{th}$.\\
(iii) muons deposit 240~MeV corresponding to the average energy
released by a vertical muon crossing a 1.2~m high water-Cherenkov tank.\\
The contribution of each particle is multiplied by the weight assigned by the
thinning algorithm.  We obtain the Cherenkov signal per unit area perpendicular to
the shower plane $C_{sp}(r)$.
For the muons, the Cherenkov signal is proportional to the track length in the
station so that: $C^{\mu} = C^{\mu}_{sp}$, whereas for the electromagnetic component:
$C^{em}= \cos\theta~C^{em}_{sp}$.

\begin{figure}
\begin{center}
\centerline{
\includegraphics[draft = false,scale = 0.7]{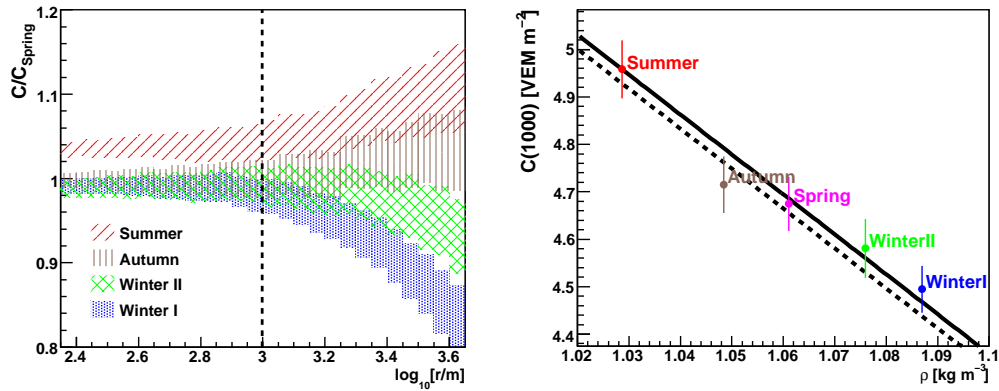}
}
\caption{
Results from the proton shower simulations with $E$~=~$10^{19}$~eV  and $\theta =
18^\circ$. Left: lateral distribution of the water Cherenkov signal per unit area
perpendicular to the shower axis $C(1000)$ in four atmospheres normalized to the
Spring one. The uncertainty is due to shower-to-shower fluctuations. Right:
$C(1000)$ as a function of $\rho$ for the five atmospheres considered. The dashed
and the continuous lines are the projections of the fit in the ($C(1000)$,~$\rho$)
plane for $P = 856$~hPa and $P = 862$~hPa, respectively.
}
\label{fig:ldf_ratio}
\end{center}
\end{figure}

The left panel of Fig.~\ref{fig:ldf_ratio} shows the lateral distribution  $C(r) =
C^{em}(r)+C^{\mu}(r)$, which is proportional to $S(1000)$, for four atmospheres
(relative to the Spring one) in the case of $E$~=~$10^{19}$~eV and $\theta =
18^\circ$. The effect related to the Moli\`ere radius can be clearly seen as a
broadening of the lateral distribution with increasing temperature. 

\begin{figure}
\begin{center}
\centerline{
\includegraphics[draft = false,scale = 0.66]{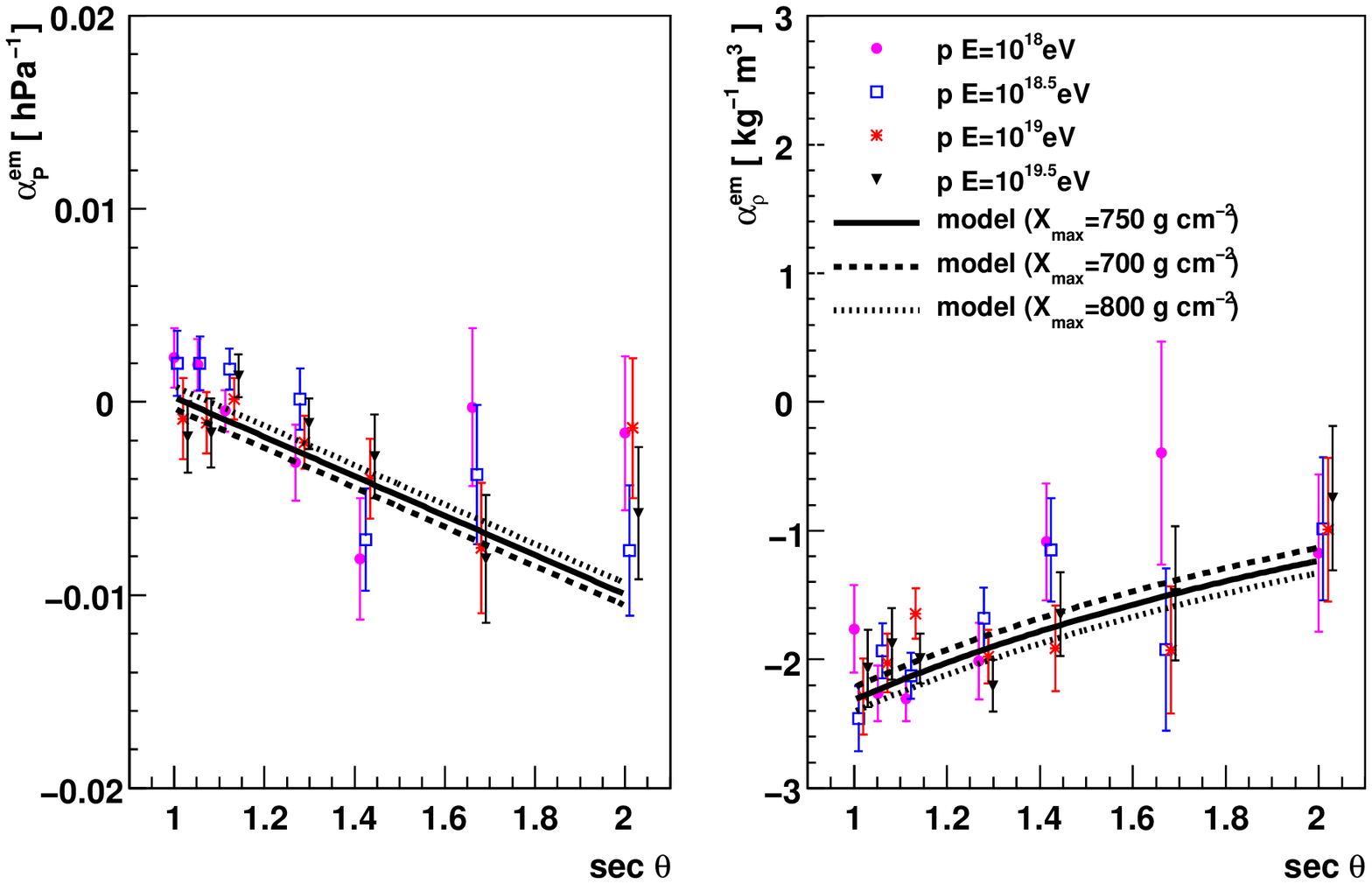}
}
\centerline{
\includegraphics[draft = false,scale = 0.66]{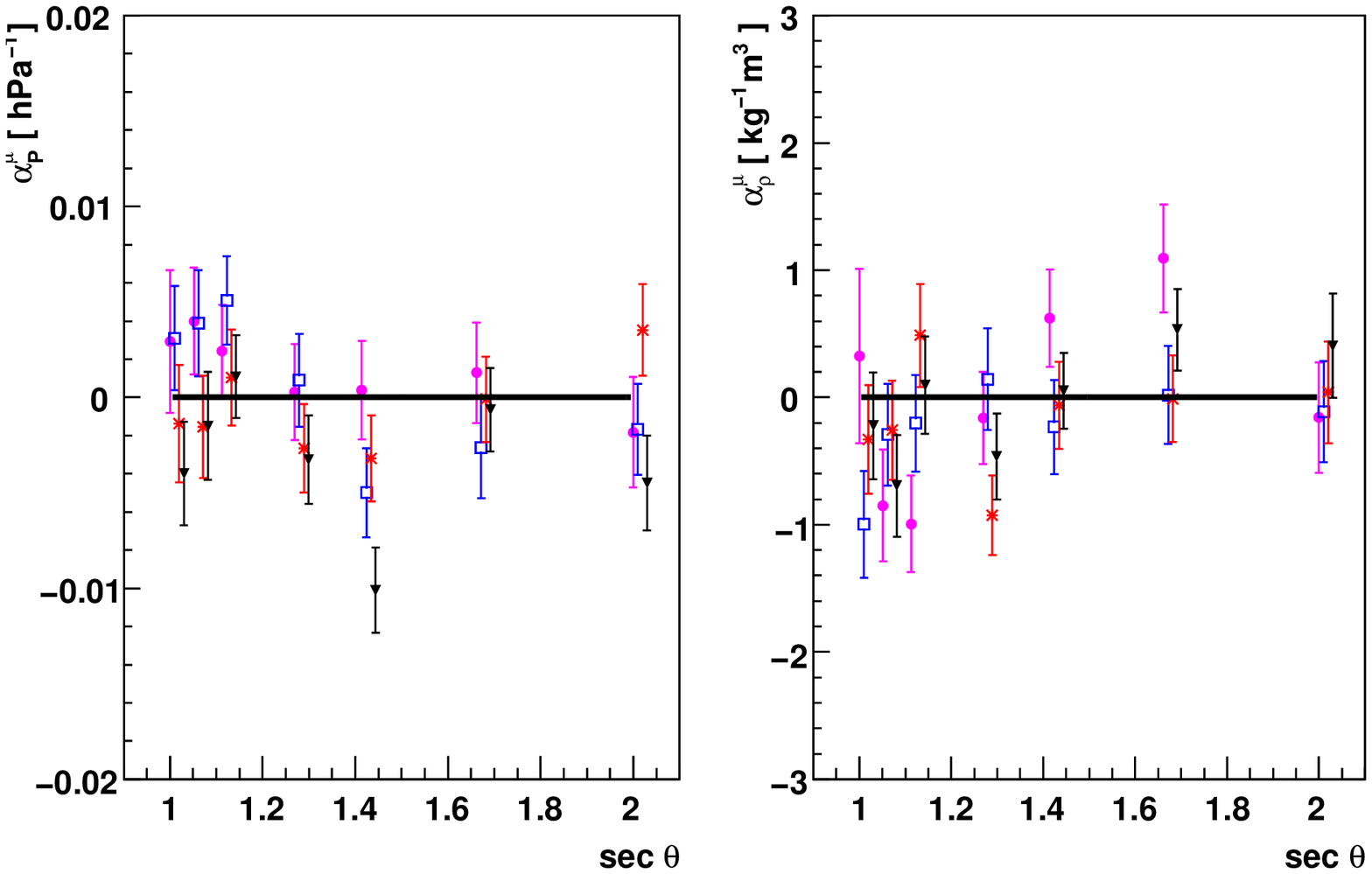}
}
\caption{
Top: atmospheric coefficients ($\alpha_P$ on the left and
$\alpha_\rho$ on the right) for the electromagnetic component as a function of
sec~$\theta$. The differently coloured markers correspond to the four
simulated energies and the lines represent the model for three different
values of $X_{max}$. Bottom: $\alpha_P$ (left) and $\alpha_\rho$ (right)
for the muonic component as a function of sec~$\theta$.}
\label{fig:coeffAllComponent}
\end{center}
\end{figure}

To derive the atmospheric coefficients, we correlate the simulated $C(1000)$
(taken as the average signal between 950~m and 1050~m) with $P$ and $\rho$ (see
eq.~\ref{eq:Sgeneral}). 
Since we are using seasonal atmospheric profiles, we do not have access to the
diurnal variation of $T$ and thus we cannot determine the coefficient $\beta_\rho$
related to the diurnal variation of $\rho$. The two coefficients $\alpha_{\rho}$
and  $\alpha_{P}$ can be determined for each fixed energy and angle with a two
dimensional fit of the $C(1000)$, obtained for the five atmospheric profiles, as
function of $\rho$ and $P$. 
As an example, we show in Fig.~\ref{fig:ldf_ratio} (right) the results of the fit
for the case of $E$~=~$10^{19}$ eV and $\theta = 18^\circ$, projected on the
($C(1000)$,~$\rho$) plane for the sake of clarity. Moreover, in the case of
simulations we are able to separate the electromagnetic and the muonic contribution to the
signal and thus to determine the atmospheric coefficients for each component (see
Fig.~\ref{fig:coeffAllComponent}).

%
\section{Comparison among model, data and simulations}
\label{sec:comparison}
In this section, we compare the atmospheric coefficients derived from data with
those expected from the model and simulations. We recall that with the simulations
we cannot access the coefficient $\beta_\rho$, as we use average seasonal profiles
for the atmosphere, while we can investigate the behaviour of separate
coefficients for the electromagnetic and muonic components of EAS. 
On the other hand, with
experimental data we cannot 
separate the electromagnetic and muonic components, while we can fully
investigate the diurnal effects of atmospheric changes and compare
measurements and expectations for all of the three coefficients.

The comparison between atmospheric coefficients for the electromagnetic and muonic components
of EAS from simulations and model is shown in Fig.~\ref{fig:coeffAllComponent}, as
a function of sec~$\theta$.  With respect to the electromagnetic part, 
the model predictions for both the $P$ and $\rho$ coefficients, and their
dependence on the shower zenith angle, are reasonable at all energies. 
Concerning the muonic component of the signal and its dependence on $P$, $\alpha^{\mu}_P$ is
compatible with zero at all energies, as expected from the flat longitudinal
development of the number of muons. For the dependence on $\rho$, the model is
not predictive but from the simulations we get a value of $\alpha^{\mu}_{\rho}$
compatible with zero.  This justifies the adoption in the model of vanishing
coefficients  for the muonic component. 

\begin{figure}
\begin{center}
\centerline{
\includegraphics[draft = false,scale = 0.5]{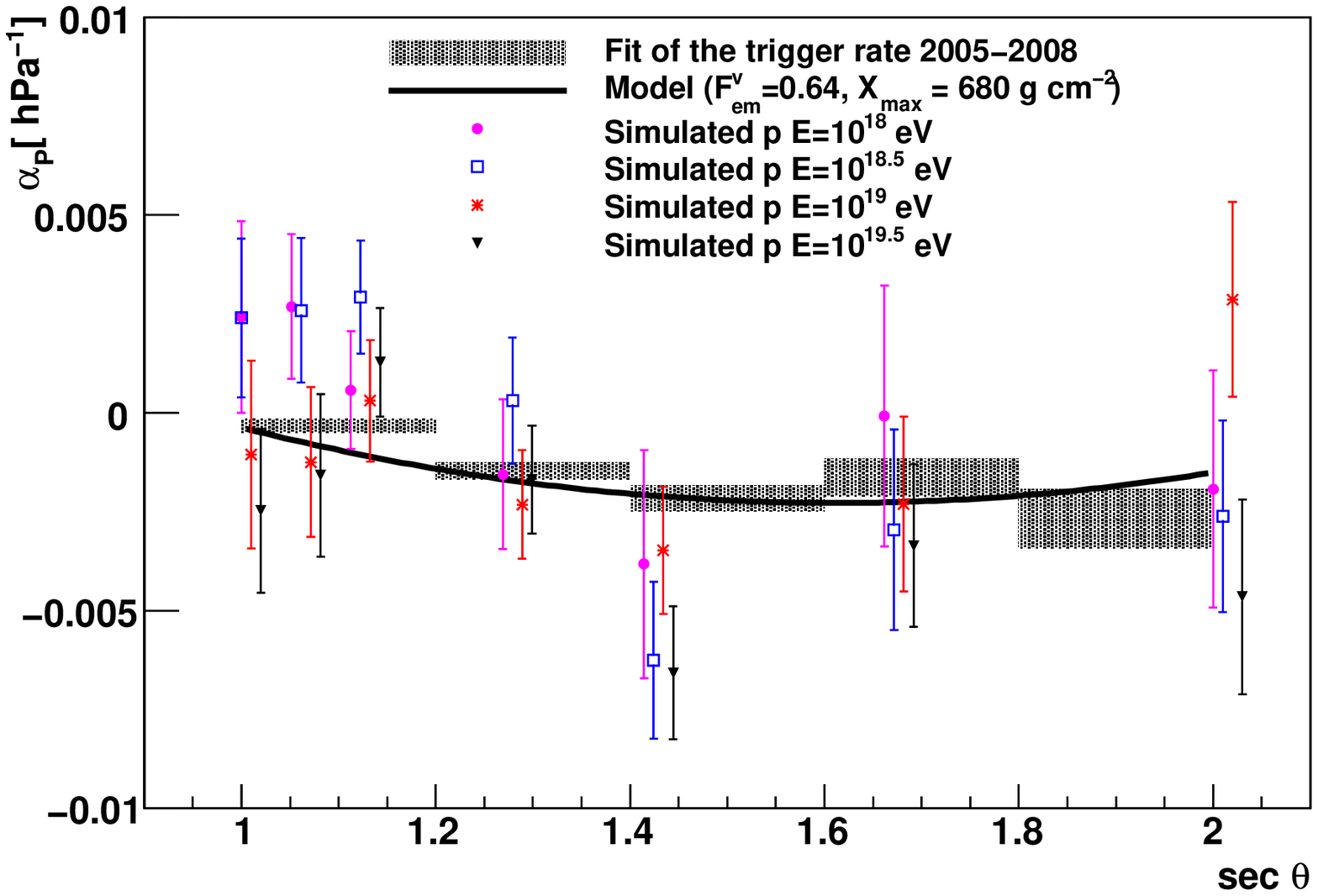}}
\centerline{
\includegraphics[draft = false,scale = 0.5]{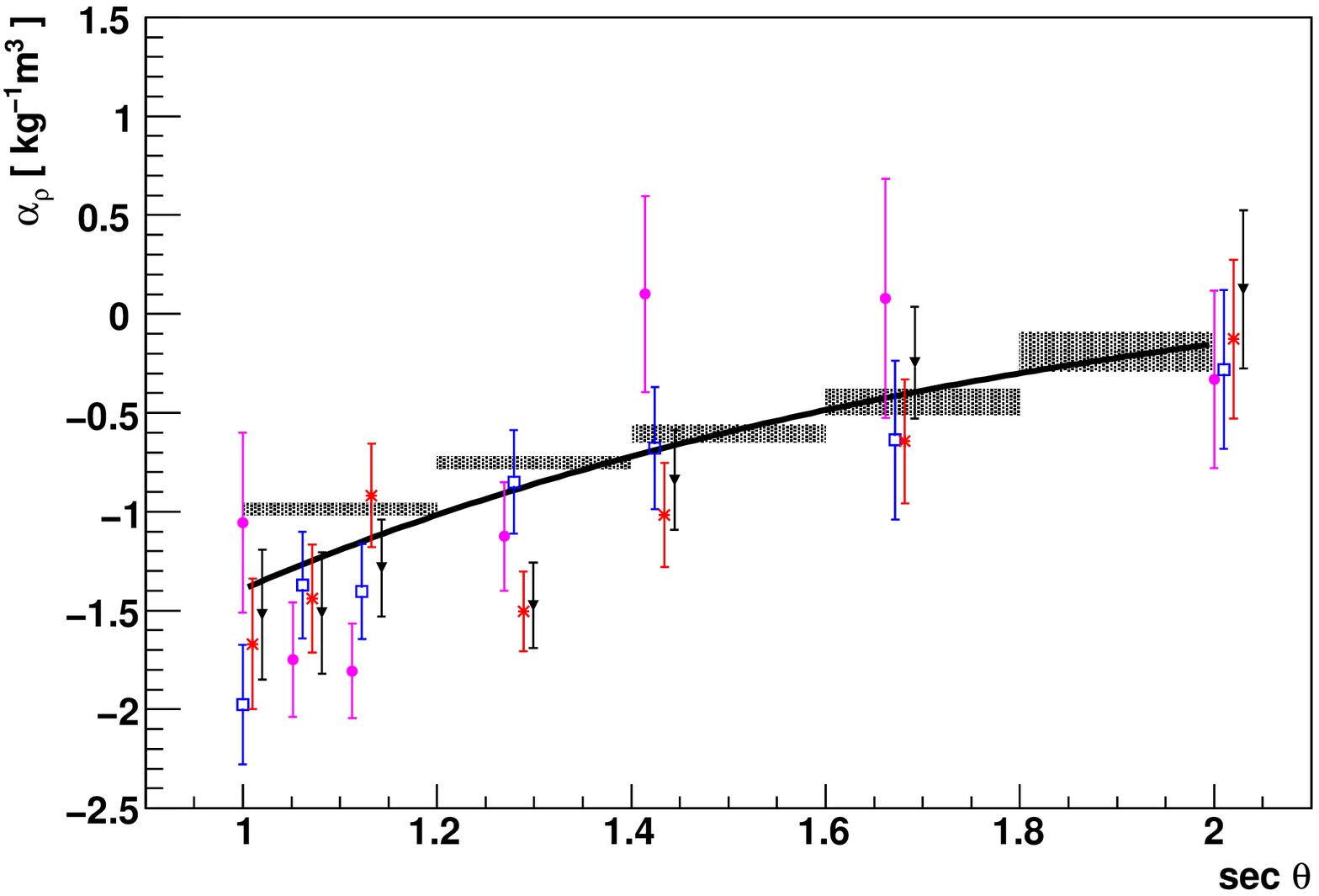}}
\caption{Comparison of the $P$ coefficients (top) and of the daily density
coefficients (bottom) as a function of sec~$\theta$ obtained from data (grey
shaded rectangle), simulations (bullets) and model (continuous line).}
\label{fig:alphacomp}
\end{center}
\end{figure}

\begin{figure}
\begin{center}
\centerline{
\includegraphics[draft = false,scale = 0.5]{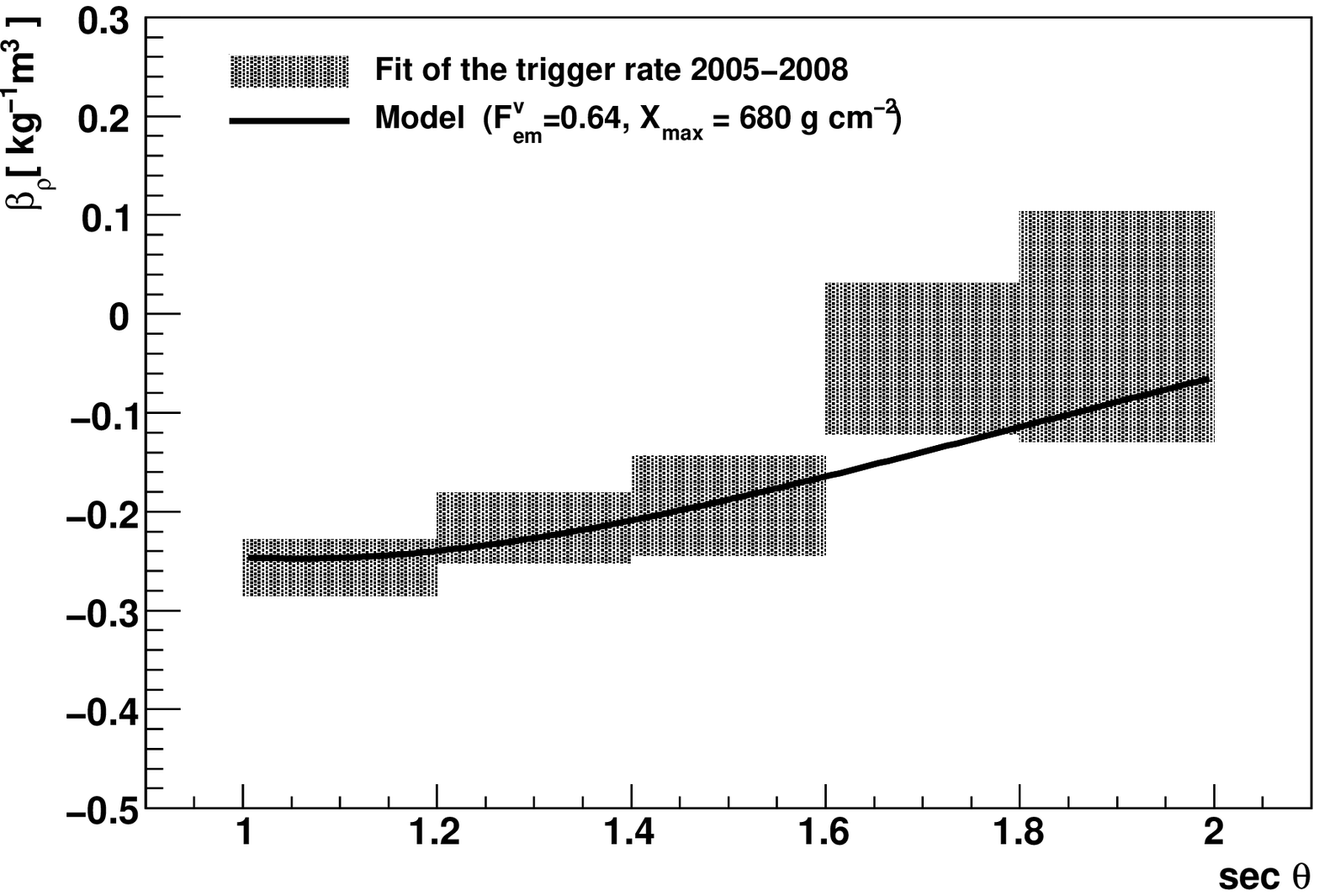}}
\caption{Comparison of $\beta_\rho$ from data with model. A fit to the data points
is performed to get the value of the parameter $a = 1.7\pm0.1$ (see
eq.~\ref{eq:betvsal}).}
\label{fig:betadcomp}
\end{center}
\end{figure}

The comparison of the global coefficients as a function of sec~$\theta$ is done
for $\alpha_P$, $\alpha_\rho$ and $\beta_\rho$  in Figs.~\ref{fig:alphacomp} and
\ref{fig:betadcomp}.
In the case of the data, the dependence on $\theta$ has been studied by dividing
the data set in subsets corresponding to five bins of equal width in sec~$\theta$.
For each subset the same fitting procedure as illustrated in
section~\ref{sec:rate} is used. The signal coefficients are then derived 
by dividing
the rate coefficients by $B(\gamma-1)$ (see the end of
section~\ref{sec:modelrate}). Since the
bulk of the triggering events have an energy $<10^{18}$~eV, we used 
$\gamma=3.30\pm0.06$, as measured  with the Auger Observatory below
$10^{18.65}$~eV \cite{icrcCombSp}.

%

The comparison among data, simulations and model is shown for the pressure
coefficient $\alpha_P$ and the daily component of the density coefficient
$\alpha_{\rho}$ in Fig.~\ref{fig:alphacomp} (top and bottom respectively). 
In the model, we use the value of $X_{max}$ measured 
by the Auger Observatory at the median energy of the triggering
events \cite{augerER}, 
and a $F_{em}^v$, corresponding at the same energy, obtained
under the assumption that $F_{em}^v$ scales linearly with the
logarithm of the primary energy. The reduced $\chi^2$ for the
data-model comparison is 3.3 for $\alpha_P$ and 11.0 for $\alpha_{\rho}$. 
For the instantaneous 
density coefficient $\beta_{\rho}$, the comparison between data and
model is shown in Fig.~\ref{fig:betadcomp}. The data-model comparison
gives in this case a reduced $\chi^2$ of 0.6.

%
%
\section{Conclusions}
\label{sec:conclusion}
We have studied the effect of atmospheric variations (in $P$, $T$ and
$\rho$) on extensive air showers using about 960$\,$000 events collected
by the surface detector of the Pierre Auger Observatory from 1 January
2005 to 31 August 2008. We observe a significant modulation of the
rate of events with the atmospheric variables, both on a seasonal scale
($\sim$ 10\%) and on a shorter time scale ($\sim$ 2\% on average during
a day).
This modulation can be explained as due to the impact of the density
and pressure changes on the shower development, 
which affects the energy estimator $S(1000)$, the size of the shower signal 
 1000~m from the shower axis. 
This affects the trigger probability
and the rate of events above a fixed energy. 

The dominant effect is due to the change with the air density of the
Moli\`ere radius near ground.
It induces a
variation of the rate of events with associated correlation coefficients
of $(-1.99\pm0.04)$~kg$^{-1}$m$^{3}$
and $(-0.53\pm0.05)$~kg$^{-1}$m$^{3}$ on long and short time scales,
respectively.

The second effect is due to the pressure changes, which affect, through
the variation of the amount of matter traversed, the stage of
development of the showers when they reach ground. The impact of the pressure variation
on the rate amounts to $(-2.7\pm0.3)\times10^{-3}$~hPa$^{-1}$.

Comparing the coefficients obtained from data, shower simulations in
different atmospheric profiles and expectations from the model built,
a good agreement is obtained, not only for the overall size of
the effect but also for the zenith angle dependence.

Taking into account the atmospheric effects will allow to reduce the
systematics in the energy reconstruction. Furthermore, it will be
possible to correct for the seasonal modulation, which can affect the
search for large scale anisotropies.

%
\section{Acknowledgements}
The successful installation and commissioning of the Pierre Auger Observatory
would not have been possible without the strong commitment and effort
from the technical and administrative staff in Malarg\"ue.

We are very grateful to the following agencies and organizations for financial support: 
Comisi\'on Nacional de Energ\'ia At\'omica, 
Fundaci\'on Antorchas,
Gobierno De La Provincia de Mendoza, 
Municipalidad de Malarg\"ue,
NDM Holdings and Valle Las Le\~nas, in gratitude for their continuing
cooperation over land access, 
Argentina; 
the Australian Research Council;
Conselho Nacional de Desenvolvimento Cient\'ifico e Tecnol\'ogico (CNPq),
Financiadora de Estudos e Projetos (FINEP),
Funda\c{c}\~ao de Amparo \`a Pesquisa do Estado de Rio de Janeiro (FAPERJ),
Funda\c{c}\~ao de Amparo \`a Pesquisa do Estado de S\~ao Paulo (FAPESP),
Minist\'erio de Ci\^{e}ncia e Tecnologia (MCT), 
Brazil;
AVCR AV0Z10100502 and AV0Z10100522,
GAAV KJB300100801 and KJB100100904,
MSMT-CR LA08016, LC527, 1M06002, and MSM0021620859, 
Czech Republic;
Centre de Calcul IN2P3/CNRS, 
Centre National de la Recherche Scientifique (CNRS),
Conseil R\'egional Ile-de-France,
D\'epartement  Physique Nucl\'eaire et Corpusculaire (PNC-IN2P3/CNRS),
D\'epartement Sciences de l'Univers (SDU-INSU/CNRS), 
France;
Bundesministerium f\"ur Bildung und Forschung (BMBF),
Deutsche Forschungsgemeinschaft (DFG),
Helmholtz-Gemeinschaft Deutscher Forschungszentren (HGF),
Finanzministerium Baden-W\"urttemberg,
Ministerium f\"ur Wissenschaft und Forschung, Nordrhein-Westfalen,
Ministerium f\"ur Wissenschaft, Forschung und Kunst, Baden-W\"urttemberg, 
Germany; 
Istituto Nazionale di Fisica Nucleare (INFN),
Ministero dell'Istruzione, dell'Universit\`a e della Ricerca (MIUR), 
Italy;
Consejo Nacional de Ciencia y Tecnolog\'ia (CONACYT), 
Mexico;
Ministerie van Onderwijs, Cultuur en Wetenschap,
Nederlandse Organisatie voor Wetenschappelijk Onderzoek (NWO),
Stichting voor Fundamenteel Onderzoek der Materie (FOM), 
Netherlands;
Ministry of Science and Higher Education,
Grant Nos. 1 P03 D 014 30, N202 090 31/0623, and PAP/218/2006, 
Poland;
Funda\c{c}\~ao para a Ci\^{e}ncia e a Tecnologia, 
Portugal;
Ministry for Higher Education, Science, and Technology,
Slovenian Research Agency, 
Slovenia;
Comunidad de Madrid, 
Consejer\'ia de Educaci\'on de la Comunidad de Castilla La Mancha, 
FEDER funds, 
Ministerio de Ciencia e Innovaci\'on,
Xunta de Galicia, 
Spain;
Science and Technology Facilities Council, 
United Kingdom;
Department of Energy, Contract No. DE-AC02-07CH11359,
National Science Foundation, Grant No. 0450696,
The Grainger Foundation 
USA; 
ALFA-EC / HELEN,
European Union 6th Framework Program,
Grant No. MEIF-CT-2005-025057, 
European Union 7th Framework Program, 
Grant No. PIEF-GA-2008-220240,
and UNESCO.

\end{document}